\documentclass[aip,graphicx]{revtex4-1}

\usepackage{siunitx}

\usepackage{graphicx}

\begin{document}

\title{Electronic structure with direct diagonalization on a D-Wave quantum annealer}

\author{Alexander Teplukhin}
\author{Brian K. Kendrick}
\email[Correspondence should be addressed to ]{bkendric@lanl.gov and pdub@lanl.gov}
\author{Sergei Tretiak}
\affiliation{Theoretical Division (T-1, MS B221), Los Alamos National Laboratory, Los Alamos, New Mexico 87545, USA}
\author{Pavel A. Dub}
\email[Correspondence should be addressed to ]{bkendric@lanl.gov and pdub@lanl.gov}
\affiliation{Chemistry Division (C-IIAC, MS K558), Los Alamos National Laboratory, Los Alamos, New Mexico 87545, USA}

\date{\today}

\begin{abstract}

Quantum chemistry is regarded to be one of the first disciplines that will be revolutionized by quantum computing. Although universal quantum computers of practical scale may be years away, various approaches are currently being pursued to solve quantum chemistry problems on near-term gate-based quantum computers and quantum annealers by developing the appropriate algorithm and software base. This work implements the general Quantum Annealer Eigensolver (QAE) algorithm to solve the molecular electronic Hamiltonian eigenvalue-eigenvector problem on a D-Wave 2000Q quantum annealer. The approach is based on the matrix formulation, efficiently uses qubit resources based on a power-of-two encoding scheme and is hardware-dominant relying on only one classically optimized parameter. We demonstrate the use of D-Wave hardware for obtaining ground and electronically excited states across a variety of small molecular systems. This approach can be adapted for use by a vast majority of electronic structure methods currently implemented in conventional quantum-chemical packages. The results of this work will encourage further development of software such as \textit{qbsolv} which has promising applications in emerging quantum information processing hardware and is able to address large and complex optimization problems intractable for classical computers.

\end{abstract}

\maketitle

\section{Introduction}

The practical difficulties in simulating many-body quantum systems such as molecules on classical computers based on central processing or graphics processing units, have been widely recognized in the quantum physics and quantum chemistry communities.\cite{review1,review2} The computational cost of the numerical solution of the time-independent electronic Schr\"{o}dinger equation in a given chemical basis set or grid scales exponentially with the number of electrons. Since the invention of digital computers in the early 1940s, the exact solution of this central quantum mechanical equation remains unfeasible for molecules having roughly more than 20 electrons.\cite{bigfci} By manipulating quantum states of matter and taking advantage of their unique features, such as superposition, entanglement and quantum tunneling, quantum computers promise to revolutionize quantum simulations of molecules and solids by bringing down the intractable cost to polynomial scaling.\cite{review1,review2} This can be achieved by using two mathematically equivalent forms of quantum computation:\cite{equiv} gate-based quantum computing and adiabatic quantum computing, which currently have prototype hardware platforms. Although gate-based quantum computers are currently available at the scale of 50-70 noisy qubits, true adiabatic quantum computers are yet to be technologically available. Existing quantum annealers (e.g. D-Wave 2000Q and upcoming Advantage devices), while naturally suited to perform adiabatic quantum computations, do not currently implement the so-called non-stoquastic Hamiltonian technology\cite{perspectives} to resolve the scaling issue.

Unsurprisingly, most of the effort to solve quantum chemistry problems on the quantum architectures has been dedicated to gate-based quantum computers. Examples include the Variational Quantum Eigensolver (VQE)\cite{vqe1,vqe2,vqe3,vqe4,vqe5,vqe6,vqe7} and the Full Quantum Eigensolver (FQE)\cite{fqe} applied to small ``toy'' molecules (e.g. H$_2$, LiH, H$_2$O, NH$_3$, etc.). Much less effort has been put toward to the use of quantum annealers. To the best of our knowledge, there are only two methods where a modern D-Wave quantum annealer was used to solve electronic structure problems to find the ground state energy.\cite{purdue,volkswagen,lumionics} These two methods could be potentially extended to the future non-stoquastic Hamiltonians. Alternatively, specifically designed techniques\cite{nonstoq} might be used instead.

The main problem in designing a method to solve the electronic structure on today's quantum annealers is to find a mapping between the electronic Hamiltonian and classical Ising model,\cite{ising1,ising2} i.e. a model of interacting spins, familiar to many physicists. The Ising model is a problem type that is accepted by today's D-Wave quantum annealers. Alternatively, one can search for a mapping to an equivalent formulation, namely Quadratic Unconstrained Binary Optimization (QUBO), which is the same as the classical Ising formulation, except that the spins are replaced with binary variables. Finding a mapping of the electronic Hamiltonian to either of these two problem types is a very challenging task.

The two recently reported mappings\cite{purdue,lumionics} were applied to small molecules, such as H$_2$ and LiH, and constitute the current state-of-the-art in solving the electronic structure on modern D-Wave quantum annealers. Both methods start with an electronic Hamiltonian in the second-quantized form and convert it to the qubit Hamiltonian using either Jordan-Wigner (JW)\cite{jw} or Bravyi-Kitaev (BK)\cite{bk} transformations. Then, they require either extensive copying of qubits\cite{purdue} or optimizing a large number of variables classically.\cite{lumionics} Moreover, both approaches have to add extra constraints (with corresponding classical strengths) and sacrifice precious qubits to lower the rank of high-order terms in their optimization problems, making those problems QUBO treatable (i.e. D-Wave compatible). Importantly, both methods were shown to work on the D-Wave 2000Q system.\cite{volkswagen,lumionics}

In our previous study\cite{qae} we reported the Quantum Annealer Eigensolver (QAE), to solve the vibrational quantum problem for diatomic and triatomic molecules. Here we generalize this method to the ubiquitous electronic quantum problem and calculate the ground and excited electronic states of a number of small molecules as a function of the nuclear positions. In contrast to the methods discussed above, the QAE is Hamiltonian and basis agnostic, uses an efficient wave function encoding scheme (powers of two instead of the extensive copying of qubits\cite{purdue}) and is quantum hardware dominant (one classical variable instead of many\cite{lumionics}). Also, the QAE does not have any conversions to pairwise form, because all terms are quadratic by construction. However, the QAE scales exponentially with the number of electrons or spin-orbitals, similar to the two previous approaches. The QAE algorithm is also redesigned in this work, reducing the computation time and increasing accuracy and usability. The D-Wave 2000Q quantum annealer is used in all representative calculations.

\section{Approach}
\label{sec:qae}

\begin{figure}
\includegraphics[scale=0.7]{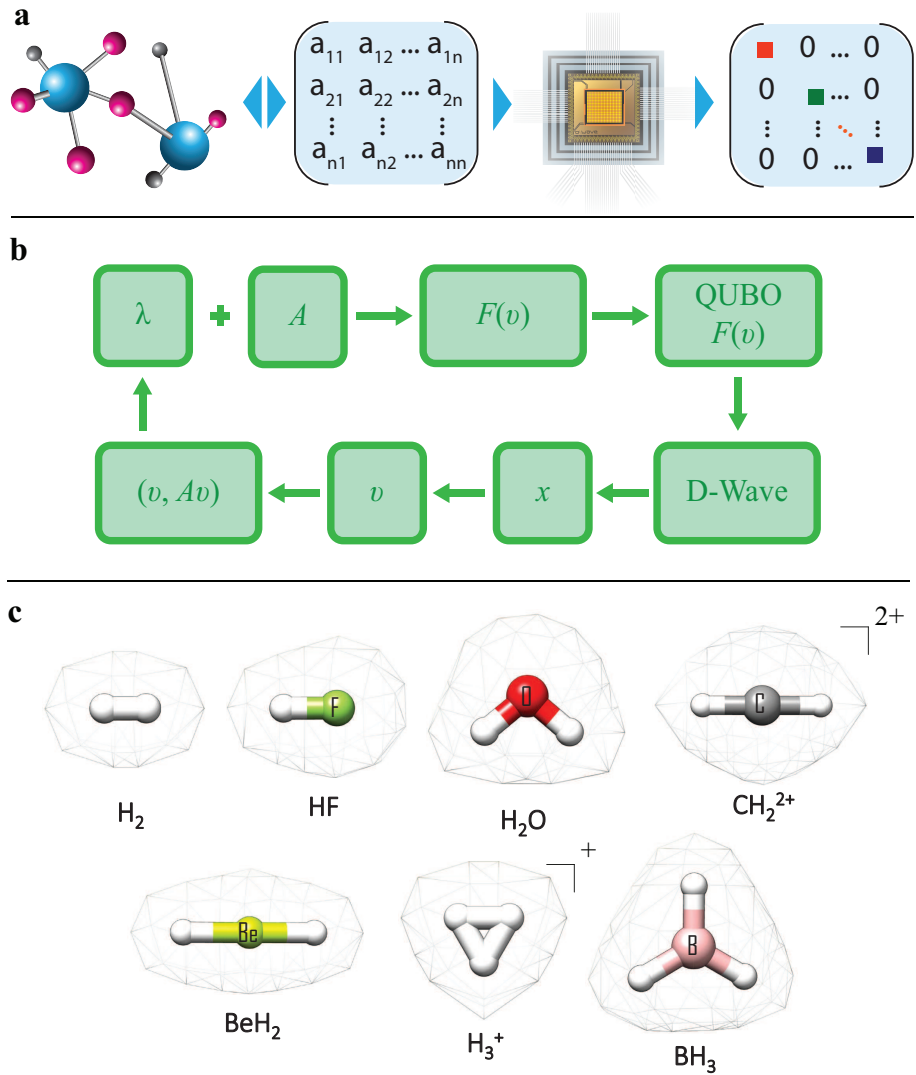}
\caption{\fontsize{10}{12}\selectfont \label{fig01} (a) Solving the electronic structure problem on a quantum annealer. Starting from Cartesian coordinates for atoms, atomic numbers and a basis set, an electronic Hamiltonian matrix is constructed in the basis of Slater Determinants of spin-orbitals by using classical quantum chemistry codes. The matrix is further diagonalized on a D-Wave 2000Q system using the QAE algorithm. (b) Diagram of the QAE iterative workflow. The objective function $F(v)$, defined by the normalization penalty $\lambda$ and the input matrix $A$, is converted to QUBO and minimized on a D-Wave quantum annealer. The optimal string of binary variables $x$ is then used to construct the vector $v$, evaluate the eigenvalue as $(v,Av)$ and guide the next choice of $\lambda$. (c) Molecular species used in the present work to demonstrate the QAE based approach. }
\end{figure}

In this section, we present a new method to solve the electronic structure using the D-Wave quantum annealer. The method has two steps, see Figure~\ref{fig01}a. In the first step, one constructs an electronic Hamiltonian matrix in the basis of Slater Determinants (SDs), which in turn, are constructed from one-electron self-consistent Hartree-Fock molecular orbitals optimized in a finite chemical basis set. In the present study, we generate Full Configuration Interaction (FCI) and Complete Active Space Self-Consistent Field (CASSCF) matrices\cite{szabo} using an in-house modified Psi4 code\cite{psi4} to account for full or a selected active space of molecular orbitals. In the second step, the QAE is used to solve the matrix for a few eigenvalues and eigenvectors. The resulting eigenvalues are the electronic energies, whereas the eigenvectors are the electronic wave functions. Since the QAE is a general-purpose method, the matrix can be constructed using any operator, basis and software.

Internally, the QAE maps the eigenvalue problem to the QUBO problem, solvable by D-Wave annealers. The mapping is based on minimization of the Rayleigh-Ritz quotient (RRQ) $R_A=(v,Av)/(v,v)$, where $A$ is an input matrix and $v$ is a vector. The minimum of the RRQ is the smallest eigenvalue, whereas the optimal vector $v$ is the corresponding eigenvector of the matrix $A$. Using an efficient power-of-two scheme, the vector $v$ is encoded in binary variables $x_i \in \{0, 1\}$ to convert the scalar product $(v,Av)$ to the QUBO expression. The norm $\|v\|=1$ is enforced by adding a constraint with a classically-optimized Lagrange multiplier $\lambda$ to the QUBO. The final form of the optimized objective function is $F(v)=(v,Av)+\lambda\cdot(v,v)$. The RRQ minimization is also known as the variational method in quantum chemistry.\cite{szabo}

Figure~\ref{fig01}b shows the overall QAE workflow. The input matrix $A$ and a $\lambda$ guess define the effective objective function $F(v)$. The function is then converted to the QUBO expression, which is minimized on the D-Wave quantum annealer, see Methods. Once an optimal binary string $x$ is obtained from the hardware, the vector $v$ is constructed using the power-of-two scheme, normalized and then used to evaluate $(v,Av)$. If none of the stopping conditions are met, then a new $\lambda$ guess is generated and the iterations continue. While iterating, all $v$ values are stored, and when the algorithm stops, the smallest $(v,Av)$ value together with the associated $v$ are returned to the user as a sought eigenpair.

The excited states are computed in a similar fashion, except that a number of eigenvalue shifts are applied to the input matrix $A$ to move the previously computed eigenvalues higher in the eigenspectrum, based on Brauer's theorem.\cite{brauer} Thus, to compute $n$ eigenpairs, the QAE performs $n$ serial runs, each time with an appropriately modified matrix $A$. 

In the original QAE implementation,\cite{qae} the normalization penalty $\lambda$ was scanned within a user-specified range with a user-specified step size. In the present study, the $\lambda$-search was completely automated, see Methods. This made the calculations faster, more accurate and easier to run, as only a matrix needs to be provided on input.

The dimensionality of the problem does not explicitly manifest anywhere in the QAE formalism, because it is part of the input matrix. If the matrix represents an operator in a direct product basis set, then the matrix dimension, the number of qubits and the QUBO size grow exponentially with the number of dimensions. We refer the reader to the original QAE paper for more details.\cite{qae} In that paper, QAE  was used to diagonalize a matrix for the molecular vibrational operator in a direct-product Fourier basis set.

As a final note in this section, a typical run of QAE requires 1) a large number of QUBO variables or qubits and 2) all-to-all connectivity between them, so that every product of two QUBO variables must be supported by the QUBO solver. These two requirements are satisfied for the classical QUBO solvers. However, they present a problem for the modern quantum annealers. The annealers have a small number of fully-connected qubits, emulated on top of a larger number of loosely-connected physical qubits. To overcome this limitation of the modern hardware, the QAE is using a QUBO software called qbsolv,\cite{qbsolv} which divides a large QUBO problem into smaller chunks (subQUBOs) and minimizes them individually. SubQUBOs are solved either classically using the Tabu search or the D-Wave quantum annealer. The Tabu search is an efficient local search technique that discourages the search from coming back to previously-visited solutions.\cite{tabu} In this way, qbsolv and, as a consequence the QAE, can be used in two modes: classical or hardware. While processing the multiple subQUBOs, qbsolv also does extra refinement steps to improve the quality of a QUBO solution.

\section{Results}

Any electronic structure problem, formulated as an eigenvalue problem, can be solved on a quantum annealer using the QAE. For example, the matrices, generated using the FCI method in the basis of Slater determinants, can be readily solved by the QAE. This will give the ground state energy and wave function, the excited states energies and wave functions and even potential energy surfaces, if the QAE is applied to multiple molecular geometries. We discuss examples of all these QAE applications for the representative molecular species shown in Figure~\ref{fig01}c.

In what follows we will be comparing the energy errors to the chemical accuracy, generally considered to be less than  1 kcal/mol. This level of accuracy of quantum chemistry methods has been christened twenty years ago and advocated by John Pople in his Nobel lecture.\cite{kcal} The main motivation behind this target is that data has to be reproduced and predicted to within experimental accuracy, being  $\sim$1 kcal/mol for energies such as heats of formation or ionization potentials. Another motivation is that at room temperature the energy change of 1 kcal/mol corresponds to an order of magnitude change in a rate constant, making this a convenient standard in kinetics studies as well.

\subsection{Ground state calculations}

The list of molecules chosen to perform proof-of-concept simulations is shown in Fig.~\ref{fig01}c. All molecular geometries were first pre-optimized at the Hartree-Fock level (gas phase) by using one of the tabulated chemical basis sets as specified below.

For the simplest H$_2$ molecule we generated multiple matrices using FCI and different basis sets. The basis set convergence is shown in Figure~\ref{fig02}a, while the full table with absolute electronic energies is given in the Supplementary Information (SI), Table~S1. In total, we used 14 basis sets, ranging from the smallest STO-3G basis set to the largest aug-cc-PVQZ basis set. The respective matrix sizes for H$_2$ range from 2x2 to 1256x1256. The number of qubits, or QUBO size, depends on the desired level of accuracy. We found that using 10 qubits per eigenvector element is sufficient, as will be discussed later. With this choice, all QUBO sizes were, therefore, 10 times larger than the matrix sizes, as shown in the third column of Table~S1. To benchmark the QAE, we diagonalized all matrices using a standard linear algebra routine (reference diagonalization via SciPy) and collected the lowest energy $E_{ref}$. As expected, larger basis sets give lower ground state energy, except for aug-cc-PVDZ, which causes a kink in the Figure. Thus for H$_2$, this basis, while being larger than all others before it, is not representative. Once we obtained the true energies from the classical SciPy calculations, we applied the QAE in both modes of operation, classical and hardware, which resulted in two energies, $E_{cl}$ and $E_{hw}$, respectively. In practice, we find that the ground state energies obtained using both modes are not very different, with the maximum difference being about 1 kcal/mol. Comparison with the reference diagonalization energies $E_{ref}$, shows that the QAE error may reach 4.8 kcal/mol for large matrices, however, it is less than 1 kcal/mol for the small matrices, namely those that have a size less than 60 (i.e., the first 6 matrices in Table~S1). The $E_{hw}-E_{ref}$ error does not exceed 0.01 kcal/mol for the first 3 smallest matrices with sizes under 10. The aug-cc-PVDZ kink is properly reproduced by the QAE. We also found that the hardware mode fails for the problems approaching 10$^4$ QUBO variables, which is probably due to the lack of memory. Because of that we could not obtain the $E_{hw}$ energy for the aug-cc-PVQZ basis set. In contrast, this particular case can be addressed with the classical mode. In addition to the problem size limit, the hardware mode runs much slower and seems to have a higher chance to fail than the classical mode (probably due to internal hardware interface library issues). To cope with the fragile and slow hardware mode, we added a checkpoint mechanism to the QAE which allows it to restart a calculation if it has stopped.

\begin{figure}
\begin{minipage}{0.5\textwidth}
\includegraphics[width=\textwidth]{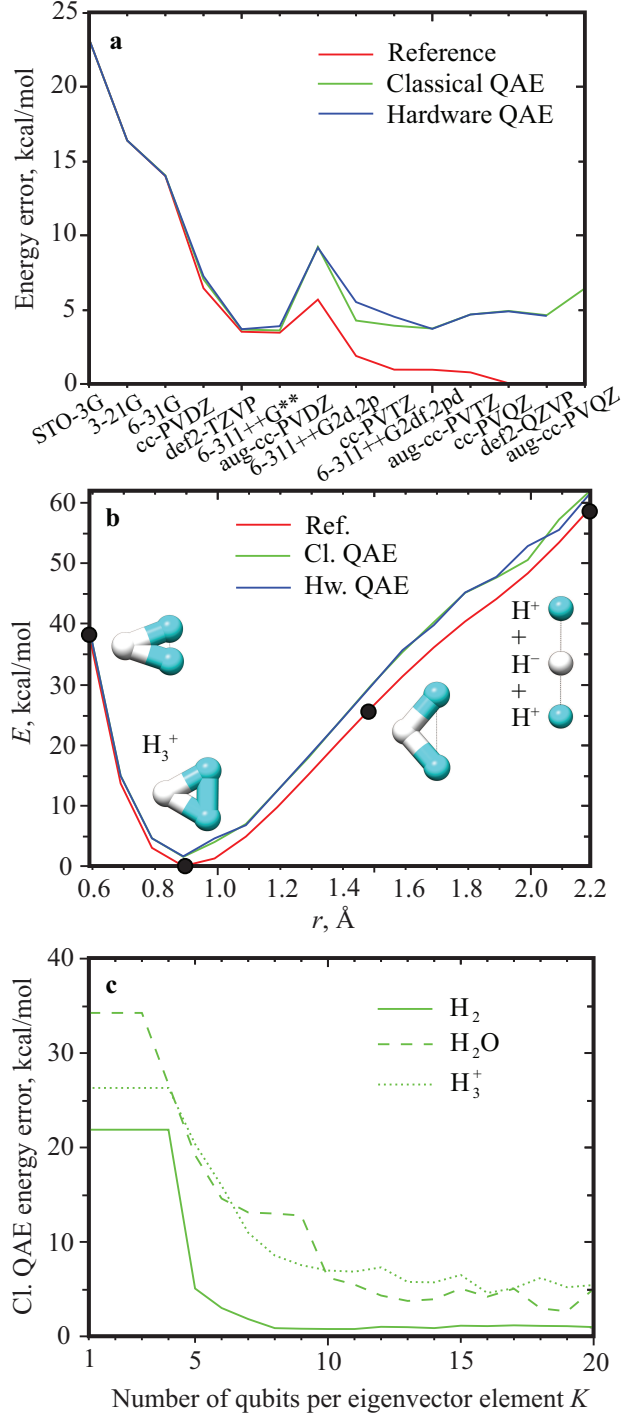}
\end{minipage}
\begin{minipage}{0.4\textwidth}
\caption{\fontsize{10}{12}\selectfont \label{fig02} (a) Convergence of the H$_2$ ground state energy with basis set. The calculations are ordered with increasing Hamiltonian matrix size. The matrices were diagonalized directly (red) and using the QAE in both classical (green) and hardware (blue) modes. The error is given relative to the energy of def2-QZVP reference diagonalization, which is the lowest among all calculations. The QAE error closely follows the reference diagonalization error for small basis sets, but exceeds it by about 5 kcal/mol for larger basis sets. (b) Potential energy curve of H$_3^+$ computed at FCI/cc-PVDZ level using the same methods as in the top panel (same colors). Molecular images (taken at black points) show how the minimum geometry evolves as a function of the distance between two terminal hydrogens. (c) Convergence of the ground state energy with respect to the number of qubits per eigenvector element $K$. The convergence study was done for H$_2$ (solid), H$_2$O (dashed) and H$_3^+$ (dotted). The QAE was running in the classical mode. Ten-qubit discretization is sufficient for all three molecules. }
\end{minipage}
\end{figure}

The electronic ground state energy errors for all molecules, studied in the present work, are collected in Table~\ref{table1}. This time we address a broader chemical space, rather than basis set expansion. For this reason, most molecules in the table were computed using the FCI method in the minimal STO-3G basis set, except a few exceptions as specified in the table. The number of qubits per eigenvector element remains 10. Similar to H$_2$, the difference between the modes $E_{cl}$ and $E_{hw}$ is less than 1 kcal/mol, except H$_2$O, for which it reaches 1.7 kcal/mol (likely due to the qbsolv as will be discussed later). The deviation from the reference energy ranges from 0.0 to 7.3 kcal/mol. Again, it was not feasible to solve a large 1256x1256 FCI problem for BH$_3$ in the hardware mode (the calculation requires a large amount of memory and more than 4 days to finish). As we have found, the hardware implementation of the QAE is limited to a matrix size of about 600, and matrices of a larger size can only be solved in the classical mode. The absolute energies are given in Table~S2, whereas the choice of active space for H$_2$O and BH$_3$ is shown in Figures~S1 and S2.

\begin{table*}
\caption{\label{table1} Electronic ground state energy errors in kcal/mol.}
\begin{ruledtabular}
\begin{tabular}{cccccccccc}

Molecule & Method & Basis & Mat. size & QUBO size & $E_{cl}-E_{hw}$\footnotemark[1] & $E_{hw}-E_{ref}$\footnotemark[2] \\

\hline

H$_2$       & FCI           & STO-3G     & 2x2       & 20   &  0.000 & 0.000 \\
HF          & FCI           & STO-3G     & 18x18     & 180  & -0.063 & 0.214 \\
H$_2$O      & FCI           & STO-3G     & 133x133   & 1330 &  0.085 & 5.399 \\
H$_2$O      & CAS(8e,7o)SCF & cc-PVDZ    & 321x321   & 3210 &  1.660 & 7.252 \\
CH$_2^{2+}$ & FCI           & STO-3G     & 169x169   & 1690 &  0.293 & 4.972 \\
BeH$_2$     & FCI           & STO-3G     & 169x169   & 1690 & -0.301 & 2.380 \\
H$_3^+$     & FCI           & cc-PVTZ    & 532x532   & 5320 & -0.071 & 7.005 \\
BH$_3$      & CAS(6e,6o)SCF & 6-311++G** & 208x208   & 2080 & -0.473 & 7.145 \\
BH$_3$      & FCI           & STO-3G     & 1250x1250 & 12500\footnotemark[3] & - & -

\end{tabular}
\end{ruledtabular}
\footnotetext[1]{Energy difference between the classical ($E_{cl}$) and hardware ($E_{hw}$) QAE modes.}
\footnotetext[2]{Energy difference between the QAE in the hardware mode ($E_{hw}$) and the reference diagonalization ($E_{ref}$).}
\footnotetext[3]{The QAE failed in the hardware mode (see text).}
\end{table*}

In summary, the chemical accuracy of 1 kcal/mol is reached for small molecules and small basis sets, which demonstrates the applicability of our QAE-based approach for the ground state calculations. However, the errors increase for larger molecules and basis sets, up to 7 kcal/mol in the present study, suggesting that the overall approach still needs to be improved. The qbsolv software likely introduces the largest errors as will be discussed in detail in the Discussion section.

\subsection{Excited state calculations}

We also calculate the lowest five singlet state energies (including the ground state) for the water molecule, labeled as S$_0$, S$_1$ ..., and report four ground-to-excited state transition energies in Table~\ref{table2}. In this set of calculations, the FCI/STO-3G matrix (133x133) was solved for eigenvalues using the reference diagonalization and the QAE in both classical and hardware modes. For the excited states, the QAE modifies the matrix after each energy is computed according to the Brauer's theorem, as described in Sec.~\ref{sec:qae}. Since a QAE eigenvector is not calculated exactly (due to limitations of qbsolv and the finite number of qubits), each spectral transformation that is based on a computed eigenvector outer product, adds a random noise to the modified matrix, causing the higher states to be calculated with less precision. This trend is seen in the last column of the table, $T_{hw}-T_{ref}$. The smaller error for the S$_0$ $\rightarrow$ S$_3$ transition (when compared to the S$_2$ and S$_4$ transitions) seems to be fortuitous, as rerunning the QAE gives somewhat different energies each time, due to the limitations of the qbsolv (see Discussion and Methods). Similarly, the energy difference between the two QAE modes for the S$_0$ $\rightarrow$ S$_1$ transition is larger than the $T_{hw}-T_{ref}$ error. These inaccuracies indicate that more runs and a less noisy implementation of the entire approach are needed to obtain reliable and representative statistics.

Notably, the QAE calculates not only the excited state energies but also the corresponding wave functions represented by the respective eigenvectors (not shown). When compared to the results of the reference diagonalization $T_{ref}$, the QAE transition energy errors $T_{hw}-T_{ref}$ are about 3\%. The absolute energies of all states can be found in Table~S3.

\begin{table*}
\caption{\label{table2} Electronic transition energies of the H$_2$O molecule computed using the QAE and FCI/STO-3G matrix in kcal/mol. }
\begin{ruledtabular}
\begin{tabular}{cccccc}

Transition & $T_{ref}$\footnotemark[1] & $T_{cl}$\footnotemark[2] & $T_{hw}$\footnotemark[3] & $T_{cl}-T_{hw}$ & $T_{hw}-T_{ref}$ \\

\hline

S$_0$ $\rightarrow$ S$_1$ & 303.056 & 300.563 & 302.125 & -1.563 & -0.930 \\
S$_0$ $\rightarrow$ S$_2$ & 369.233 & 373.585 & 379.837 & -6.252 & 10.603 \\
S$_0$ $\rightarrow$ S$_3$ & 441.058 & 437.217 & 444.802 & -7.585 &  3.744 \\
S$_0$ $\rightarrow$ S$_4$ & 590.407 & 606.617 & 612.352 & -5.735 & 21.945 \\

\end{tabular}
\end{ruledtabular}
\footnotetext[1]{Transition energy obtained using the reference diagonalization.}
\footnotetext[2]{Transition energy obtained using the QAE in the classical mode.}
\footnotetext[3]{Transition energy obtained using the QAE in the hardware mode.}
\end{table*}

\subsection{Potential energy curves}

To test the QAE capacity to generate potential energy surfaces we calculate the potential energy curve of the H$_3^+$ system, as shown in Figure~\ref{fig02}b. The FCI/cc-PVDZ matrices were generated for 17 distinct geometries, each pre-optimized at the Hartree-Fock level, where the distance between the two terminal hydrogens was varied from 0.59 to 2.19 \si{\angstrom} following a step size of 0.1 \si{\angstrom} (i.e. constrained optimization). The errors behave similarly to those reported in Table~\ref{table1}. There is not much difference between the two QAE modes. The difference with the reference energy varies from 1.1 kcal/mol at 0.59 \si{\angstrom} to 4.5 kcal/mol at 1.99 \si{\angstrom}. All of the matrices were 153x153, except for the last two geometries near 2.19 \si{\angstrom} which were 51x51 due to the increasing symmetry of the system.

As with the ground and excited states calculations reported above, the QAE is shown to work for potential energy surfaces as well. However, the method results in a larger 5 kcal/mol error along the H$_3^+$ dissociation curve compared to the reference simulations.

\subsection{Number of qubits}

The number of qubits per eigenvector element $K$ is an important and the only QAE convergence parameter. This parameter controls how accurately an eigenvector is represented, when the eigenvalue problem is mapped to the QUBO problem. For example, $K=10$ translates to $2^{-10} \approx 10^{-3}$ error in the eigenvector elements, whereas $K=20$ to $10^{-6}$ error. In order to quantify its influence, we performed a number of calculations with different $K$.

The ground state energy convergence with respect to $K$ is shown in Figure~\ref{fig02}c for three molecules, H$_2$, H$_2$O and H$_3^+$. The three matrices for these cases were generated using the FCI and cc-PVDZ, STO-3G, and cc-PVTZ basis sets, respectfully. With this choice, the matrices of varying sizes are covered:  22x22 (small), 133x133 (medium) and 532x532 (large). As one can see, increasing the number of qubits improves the resolution of the eigenvector representation and lowers the ground state energy. At the same time, a plateau is reached after $K=10$ qubits and using more qubits is not necessary. Based on our tests, the noisy plateau is due to the limitations of qbsolv, see Discussion section. The ground state energy has converged to 1.0 kcal/mol for H$_2$, 4.2 kcal/mol for H$_2$O and 5.9 kcal/mol for H$_3^+$ (averaged over $K$ from 11 to 20). The plot was generated using the QAE running in the classical mode, because it is much faster than the hardware mode (also the energy difference between the two modes is smaller than the errors from qbsolv). The H$_2$O matrix is the same matrix that was used to compute the ground and excited states of water, whereas the H$_3^+$ matrix corresponds to the minimum geometry matrix in Figure~\ref{fig02}b. Ideally, this convergence study should be done for every specific case, however, verifying convergence for the three typical matrix sizes should suffice.

\section{Discussion and overview}
\label{sec:disc}

Two overarching statements can be derived from the results reported in the previous section. The first is that the QAE-based approach to the electronic structure problems is viable. The second is that the energy errors are too large (more than 1 kcal/mol), which places the QAE not even close to the capacity of the modern quantum chemistry software. The latter motivated us to examine the entire procedure, aiming at identifying a component that is responsible for the inaccurate energies. Ultimately, we found that the qbsolv software,\cite{qbsolv} while being a great tool to manage large QUBO problems, is the weak spot. This QUBO solver does not solve QUBO problems exactly. Some amount of classical ``noise'' is always present in the optimal QUBO solutions returned by the qbsolv, inadvertently ``polluting'' the QAE iterations and causing inaccuracy in the resulting electronic energies. In this section, we first gather statistics on how large a typical energy spread is, then we demonstrate the fluctuating behavior of the qbsolv and discuss possible ways to resolve this issue. Additionally, as we stated earlier, running the QAE on an actual hardware is significantly slower than running it in the classical mode. We outline possible solutions to this performance issue as well.

The energy errors reported in the present study are small and large for small and large matrices, respectfully. To quantify the degree of energy fluctuation, we perform 10 QAE runs for two molecules H$_2$ and H$_2$O. For a small 8x8 H$_2$ FCI/6-31G matrix (80 QUBO variables) the average energy error was 0.004 kcal/mol, when compared to the reference energy  $E_{ref}$, while the spread ($\sigma$ the standard deviation) was 0.005 kcal/mol. For a larger case, a 133x133 H$_2$O FCI/STO-3G matrix (1330 QUBO variables) the average error was 6.79 kcal/mol and the standard deviation was 0.47 kcal/mol.

As stated above, we found that qbsolv does not give a reproducible QUBO solution. In order to demonstrate that, we fixed the normalization penalty at $\lambda=-100$ and called the qbsolv 10 times to minimize the same QUBO. For a simple 3x3 matrix (30 QUBO variables), the qbsolv returned 10 identical solutions. For a 9x9 matrix (90 QUBO variables), only 7 solutions were the same, whereas the other 3 were unique. For the H$_2$O 133x133 matrix all 10 QUBO solutions were different with the ground state energy spread (standard deviation) of 18.7 kcal/mol (for a fixed $\lambda=-100$, which is not too far from $\lambda_{opt}=-83.9$). The fluctuating behavior of qbsolv makes it hard to obtain a reliable ground state energy and subsequently to design a robust technique for finding the optimal normalization penalty $\lambda$. On the other hand, exploring the space of $2^{1330}$ possible configurations, or even of $2^{5320} \approx 3 \times 10^{1601}$ (H$_3^+$ FCI/cc-PVTZ, the largest QUBO solved using a D-Wave annealer in the present work) is a very challenging task.

There are a number of ways to improve the accuracy of the current implementation. First of all, one can simply run the qbsolv multiple times and always choose the solution with the smallest QUBO value. Alternatively, some of qbsolv's default parameters can be adjusted. The current version of qbsolv has only one parameter that the user can specify, namely, the number of repeats (iterations) of the outer loop before the algorithm stops. Our test for a typical QAE calculation (H$_2$O 133x133 matrix) showed that the default number of repeats 50 is sufficient. However, there are other hard-coded parameters in the qbsolv code that may need to be tested as well. Both approaches (multiple qbsolv runs or tightening qbsolv internal parameters) may significantly increase the time-to-solution duration. Second, one may consider using an exact brute-force approach to minimize a QUBO (while somehow interleaving this with the calls to an annealer). An exact QUBO solver is already implemented by D-Wave and ``becomes slow for problems with 18 or more variables'',\cite{ocean} which makes it useful for testing and debugging code only. Alternatively, D-Wave started providing a new Hybrid Solver Service (HSS)\cite{hss} via their Leap cloud service,\cite{leap} which is a combination of classical and hardware solvers working together on QUBO problems up to 10,000 variables. The HSS may help decrease the QAE errors, taking into account that D-Wave strongly recommends using their dwave-hybrid framework (which seems to be the main component of the HSS\cite{hss}) instead of qbsolv.\cite{ocean} At the same time, the original qbsolv was recently upgraded and has started a new life under the ``QCI qbsolv'' name,\cite{qci-qbsolv} so it is also worth exploring if this newer version will work better with the QAE.

Running the QAE on an actual hardware has its own issues. Currently, the hardware mode is much slower than the classical mode, which might be due to several reasons. One reason is that every time the qbsolv submits a subQUBO, a mapping between subQUBO variables and the actual physical qubits, called minor-embedding, is recalculated. This takes time, since it presents a separate challenging problem. There is an option to use a lazy embedding, where the mapping computed for the very first submission will be used for all other submission, thus saving some time. We are exploring this possibility. Another piece of the intermediate software, that seems to consume time, is the integrated post-processing system of the Virtual Full-Yield Chimera (VFYC) wrapper of the D-Wave annealer. The main idea behind VFYC solvers is to classically ``fix'' unrepresented variables in the actual hardware, so that the user will always work with the same number of fully-connected qubits (full-yield), thus realizing a ``portable'' code. In practice, this increases the time-to-solution lag and may decrease the quality of subQUBO solutions. Multiple users working on a single D-Wave device introduce one more problem, when compared to the purely classical way of running the QAE and qbsolv. The interested reader is encouraged to read the discussion section of our earlier QAE paper\cite{qae} to see how the QAE may be improved. Thus, while the adiabatic quantum annealing is a viable quantum computing paradigm, it clearly has its own well-defined set of challenges.\cite{perspectives}

In summary, a novel methodology was presented for solving electronic structure problems on a quantum annealer by numerically exact diagonalization. The diagonalization was performed using the Quantum Annealer Eigensolver (QAE) methodology. Several molecules were studied spanning a wide range of complexity from diatomic to tetraatomic systems. The electronic ground and excited state energies (and their wave functions) were computed on a D-Wave quantum annealer. Diagonalizing the electronic structure matrix at multiple molecular geometries was also demonstrated and can be used to generate potential energy surfaces. As an example, we calculated the potential energy curve of the H$_3^+$ cation. Importantly, the QAE computed potential energy is a smooth function of the nuclear coordinates. The QAE gives accurate molecular energies for small molecules and matrices (energy errors are less than 1 kcal/mol), however, the errors tend to increase and fluctuate for larger matrices. We found that this is primarily due to the noisy output of the underlying QUBO solver called qbsolv, the main goal of which is to reduce a large QUBO problem to a number of small subQUBO tasks (that fit on the D-Wave quantum annealer). Although the fundamentals of the QAE did not change since we applied it to the vibrational problem\cite{qae}, in this work we generalized the normalization penalty search, which made the method faster and does not require any \textit{a priori} knowledge about an eigenvalue problem from the user.  In contrast to the other works on this subject, our method is general –- the only requirement is to formulate the problem as an eigenvalue problem. Also, the power-of-two scheme allows for the efficient use precious qubits, and having a single classically-optimized parameter $\lambda$ makes our method hardware dominant.

The electronic structure methods used in the current work are only a small representative subset of methods that can be used to generate the input matrices for the QAE. The solution of the electronic structure problem on a classical computer is ultimately mapped into a diagonalization of a Hermitian Hamiltonian matrix for a variety of approaches of different accuracy ranging from Hartree-Fock (HF) to Density-Functional Theory (DFT) to Coupled Cluster (CC) and to Full Configuration Interaction (FCI) techniques. Due to the eigenproblem representation, solving all these tasks on the upcoming quantum computing hardware is an important and natural challenge for the QAE and similar algorithms. These are just a few examples of possible typical applications of the QAE in the field of quantum chemistry. We also expect that future technological improvements to the quantum annealers will continue to increase the number of qubits and reduce the noise. The results of this work encourage further improvements to the qbsolv software, which is commonly used in the quantum annealing community to address large optimization problems. All these improvements (both hardware and software) will enable practical solutions of larger problems with increased accuracy and may ultimately challenge the capabilities of conventional computing.

\section{Methods}
\label{sec:methods}

\subsection{Generation of matrices}

All molecular geometries were first optimized at the restricted  Hartree-Fock level (gas) with the Gaussian 09 (E01) code,\cite{gaussian09e01} ensuring the highest point group symmetries. For H$_3^+$ in the cc-PVTZ basis, the optimized geometry has C$_{2v}$ point group symmetry. The restricted FCI and CASSCF matrices were then generated using an in-house modified Psi4 code\cite{psi4} with the optimal Cartesian coordinates of atoms as input data. The matrices were generated imposing the unitary groups U(1) and SU(2) (spin and particle conservation) and point group symmetries and contain nonzero matrix elements only. For Abelian point groups (e.g. C$_{2v}$), Psi4 automatically detects the molecular symmetry. If the point group is non-Abelian (e.g. D$_{\infty h}$, C$_{\infty v}$, D$_{3h}$), the detected symmetry corresponds to one of its subgroups. An energy threshold of 10$^{-8}$ Hartree for diagonal and of 10$^{-10}$ Hartree for off-diagonal elements was used. The nuclear repulsion (NN) term was manually added to all constructed matrices.

\subsection{Quantum Annealer Eigensolver}

The Quantum Annealer Eigensolver (QAE) is a general-purpose variational method to compute eigenvalues and eigenvectors of a given matrix. It is based on the min-max (or variational) theorem, that states that the smallest (largest) eigenvalue of a $n \times n$ Hermitian matrix $A$ is the minimum (maximum) of the Rayleigh-Ritz quotient $R_A=(v,Av)/(v,v)$, where $v$ is a vector. If the vector $v$ is normalized, i.e. its norm $(v,v)=\|v\|$ is 1, then $R_A=(v,Av)$. The vector $v$ that realizes the minimum (maximum), is the eigenvector associated with that eigenvalue. For the QAE, we are interested in the minimum only. Notably, $R_A(v)$ is functionally similar to the polynomials used in the Quadratic Unconstrained Binary Optimization (QUBO) problems solved by D-Wave quantum annealers. A QUBO problem is the problem of minimizing a quadratic polynomial $x^TQx$ over a vector of binary variables $x=(x_1,x_2,...,x_m)$ for a given $m \times m$ matrix of real weights $Q$. Although looking similar, $R_A(v)=\sum_{\alpha,\beta}^{n,n} v_{\alpha} A_{ij} v_{\beta}$ and $x^TQx=\sum_{i,j}^{m,m} x_i Q_{ij} x_j$ operate on different kinds of variables: the elements of the normalized vector $v$ are continuous real numbers of magnitude 1, $v_\alpha \in [-1;1]$, whereas the elements of $x$ are discrete, $x_i \in \{0, 1\}$. The QAE provides a mapping between $v$ and $x$ using a powers-of-two encoding scheme. In this scheme each element of $v$ is encoded using $K$ binary variables or qubits $q^\alpha_i$, $1 \leq i \leq K$. Each qubit is multiplied by the weight of an appropriate power of two, and one more qubit is responsible for the sign of $v_\alpha$. This encoding scheme is known as the fixed-point number representation and can represent fractional numbers between $-1$ and $+1$ up to $2^{-K}$ accuracy. The sum $x^TQx$ includes both $Q_{i,j}$ and $Q_{j,i}$ terms which in general can be non-symmetric. However, QUBO problems are by definition strictly symmetric and therefore only the $i\le j$ terms are needed. This requires a symmetric $Q$ matrix and all pairwise terms with $i<j$ are doubled in constructing the QUBO functional.

Although the mapping describe above looks natural and promising, it lacks one important component, the vector normalization. Indeed, a trivial solution to the QUBO function constructed as described, is the null vector $\mathbf{0}$ (i.e., the unique vector having zero length). To fix that, the QUBO must be augmented with a normalization constraint that ensures $\|v\|$ is unity or close to unity. As with any constraint in combinatorial optimization, the normalization condition comes with its own penalty or Lagrange multiplier $\lambda$. Since the norm can be larger or smaller than 1, a natural choice for the constraint is the quadratic form $\lambda(\|v\|-1)^2$. However, this form is biquadratic in $v$ or $x$, which is not a valid QUBO form. The workaround is to decrease the degree of the constraint to the first power, $\lambda(\|v\|-1)$, which becomes  $\lambda\|v\|$ after we drop the constant shift $\lambda$. Thus, the final QUBO function is $F(v)=(v,Av)+\lambda\cdot(v,v)$ or $F(x)=x^TQx+\lambda\cdot(x,x)$.

The optimal normalization penalty $\lambda_{opt}$ can be found in many ways. In our previous study,\cite{qae} we implemented a simple scanning in $\lambda$, where for each $\lambda$ a non-trivial solution $v_\lambda$ with the smallest $R(v)$ value was considered as the best solution. The best of the best $v_\lambda$ solutions was the final answer. However, this method of $\lambda$-searching is inconvenient as it requires a good guess for the $\lambda$ search range and also depends on a step size ($\Delta\lambda$), both of which are problem dependent.

In the present study, we made the $\lambda$-search completely automatic and iterative. Based on our experience with the scanning technique,\cite{qae} we noticed that the solution $v_\lambda$ that minimizes $F(v)$ is a zero vector $\mathbf{0}$ for a large positive $\lambda$ and is a non-zero vector for a large negative $\lambda$. The optimal $\lambda_{opt}$ value is located somewhere in between, usually around a ``phase-transition'' point (i.e., on the boundary between trivial and non-trivial solution areas). Therefore, we first construct a wide $\lambda$-range, where the best solution at the left end is non-trivial and the solution at the right end is trivial. The matrix element with the largest magnitude serves as a good guess for the range ends (multiplied by $-1$ for the left end). If a solution of the required type is not found for an end, then the guess is doubled. Once the range in $\lambda$ is determined, QAE starts iteratively shrinking it using a bisection method, such that the best solutions at the range ends do not change their type. In this way, the range reduces to a small region around the optimal $\lambda_{opt}$. Currently, we have two stopping criteria for this iterative scheme: either the change in the expectation value $R(v_\lambda)$, computed classically from optimal $x$, becomes smaller than a user specified tolerance, or the $\lambda$-range shrinks to a single point $\lambda_{opt}$. While the first criterion seems to be sufficient, it may never be triggered for inaccurate QUBO solvers (either classical or hardware). The fluctuating output of these solvers leads to a situation where the tolerance can never be reached, causing an infinite loop. Additionally, inaccurate QUBO solvers make some types of searching techniques inapplicable, such as those based on the dependence of $R(v_\lambda)$ on $\lambda$. The randomness in the vicinity of $\lambda_{opt}$ will cause gradient-based methods to fail. In this case, extrapolating techniques, such as Direct Inversion in the Iterative Subspace (DIIS)\cite{diis1,diis2,diis3} based on previous $\lambda$-points, could be of help. The overall workflow is shown in Figure~\ref{fig01}b of the main text.

The eigenpairs with larger eigenvalues are computed by applying a spectrum transformation to the initial matrix and repeating the iterative procedure described above. Specifically, based on Brauer's theorem,\cite{brauer} if $v_0$ is the eigenvector with the smallest eigenvalue, then one can construct a new matrix $A'=A+\mu_0(v_0 \otimes v_0)$, where $\otimes$ denotes the outer product operation and multiplier $\mu_0$ controls how much higher in the spectrum we want to relocate the computed eigenpair. If it was relocated high enough, then repeating the QAE iterations for $A'$ will give the next eigenpair. Thus, to compute $n$ eigenpairs, the QAE performs $n$ serial runs, each time with an appropriately modified matrix $A'$. The multipliers $\mu_i$ ($i=0 \dots n-1$) could be quite arbitrary, but each has to be larger than all previously computed eigenvalues. Currently, all $\mu_i$ shifts are set equal to the largest matrix element multiplied by 16, however, multiplying by 8, 4 or 2 gives essentially the same results, with differences smaller than other errors. Alternatively, the difference between the maximum and minimum elements of $A$, or any other dynamic range estimate, can be used as a shift. In the present work, the calculation of multiple eigenpairs is not the main focus, as only a few eigenpairs are of interest. However, the QAE can be potentially modified to find ``batches'' of eigenpairs. The shifting may need a revision or replaced with a smarter technique.\cite{saad1,saad2} Linear algebra tricks, such as a correlation of the minimum eigenvector with the matrix rows,\cite{minvec1,minvec2} could also be leveraged to aid the QAE.

\subsection{Software}

In the present work, the QAE was implemented in Python 3 and requires D-Wave's Ocean tools to be installed. The access to a D-Wave quantum annealer has to be configured separately. After the QAE has mapped an eigenvalue problem (with a chosen normalization penalty $\lambda$) to a QUBO form, the code makes a call to the qbsolv to minimize the QUBO and obtain the best QUBO solution. The qbsolv is included in the Ocean tools, and is capable of running in both classical and hardware modes. For the reference diagonalization, we used SciPy eig() function, which is a wrapper for the geev() family of functions in LAPACK.\cite{lapack}

\subsection{Quantum annealer}

LANL's D-Wave 2000Q was used in all hardware-mode calculations. This machine operates at $T=0.015K$ and has 2048 qubits and 6,016 couplers. In an actual quantum annealer some qubits and couplers are not active (unrepresented), so the total number of them is a bit smaller. One will need either to find a specific mapping between QUBO variables and physical qubits (minor-embedding) for a given machine or use a VFYC version of a hardware QUBO solver, which postprocess a QUBO solution to fix unrepresented qubits and couplers. Another issue is a connectivity between qubits. In the D-Wave 2000Q, each qubit is connected to 6 neighbors only. To increase the degree of connectivity, the minor-embedding procedure generates a chain of physical qubits to represent one logical qubit. Chains overcome the 6-neighbor limit while sacrificing physical qubits. As a result, the size of the final full graph is only 64 logical qubits. The qbsolv constructs subQUBOs of this particular size.

\section*{Code availability}

The QAE code is available from the authors upon request.

\bibliographystyle{naturemag}
\bibliography{qae}

\begin{acknowledgments}

Research presented in this article was supported by the Laboratory Directed Research and Development (LDRD) program of Los Alamos National Laboratory (LANL) under project number 20200056DR. This work was conducted in part at the Center for Integrated Nanotechnologies, a U.S. Department of Energy, Office of Basic Energy Sciences  user facility. The authors would like to thank Yu Zhang for modifying the Psi4 code to suit the project needs, as well as Susan M. Mniszewski, Petr M. Anisimov and Christian F. A. Negre for fruitful discussions.

\end{acknowledgments}

\section*{Author contributions}

A.T. developed the algorithm, performed the numerical calculations, analysis and writing of the manuscript. B.K.K. contributed to the development of the algorithm, analysis and writing of the manuscript. P.A.D. selected and optimized molecules/ions with Gaussian code and generated the input matrices with Psi4. P.A.D. and S.T. contributed to the analysis and writing of the manuscript.

\section*{Competing interests}

The authors declare no competing interests.

\end{document}

\title{Supplementary Information \vspace{0.5cm} \\ Electronic structure with direct diagonalization on a D-Wave quantum annealer}

\author{Alexander Teplukhin}
\author{Brian K. Kendrick}
\email[Correspondence should be addressed to ]{bkendric@lanl.gov and pdub@lanl.gov}
\author{Sergei Tretiak}
\affiliation{Theoretical Division (T-1, MS B221), Los Alamos National Laboratory, Los Alamos, New Mexico 87545, USA}
\author{Pavel A. Dub}
\email[Correspondence should be addressed to ]{bkendric@lanl.gov and pdub@lanl.gov}
\affiliation{Chemistry Division (C-IIAC, MS K558), Los Alamos National Laboratory, Los Alamos, New Mexico 87545, USA}

\date{\today}

\maketitle

\renewcommand{\thetable}{S\arabic{table}}

\begin{table*}
\caption{\label{table1} Electronic ground state energy of the H$_2$ molecule computed using the QAE and FCI matrices. The absolute energies are given in Hartrees ($E_\textrm{h}$), the differences are given in kcal/mol. }
\begin{ruledtabular}
\begin{tabular}{cccccccc}

Basis & Mat. size & QUBO size & $E_{ref}$\footnotemark[1] & $E_{cl}$\footnotemark[2] & $E_{hw}$\footnotemark[3] & $E_{cl}-E_{hw}$ & $E_{hw}-E_{ref}$ \\

\hline

STO-3G & 2x2 & 20 & -1.13684890 & -1.13684890 & -1.13684890 & 0.000 & 0.000 \\
3-21G & 8x8 & 80 & -1.14772560 & -1.14772396 & -1.14772157 & -0.001 & 0.003 \\
6-31G & 8x8 & 80 & -1.15152414 & -1.15145953 & -1.15151476 & 0.035 & 0.006 \\
cc-PVDZ & 22x22 & 220 & -1.16356044 & -1.16260821 & -1.16226936 & -0.213 & 0.810 \\
def2-TZVP & 36x36 & 360 & -1.16822660 & -1.16798881 & -1.16795448 & -0.022 & 0.171 \\
6-311++G** & 54x54 & 540 & -1.16832827 & -1.16808739 & -1.16763178 & -0.286 & 0.437 \\
aug-cc-PVDZ & 66x66 & 660 & -1.16477801 & -1.15912926 & -1.15921477 & 0.054 & 3.491 \\
6-311++G2d,2p & 88x88 & 880 & -1.17082208 & -1.16702940 & -1.16504499 & -1.245 & 3.625 \\
cc-PVTZ & 136x136 & 1360 & -1.17229056 & -1.16757762 & -1.16662200 & -0.600 & 3.557 \\
6-311++G2df,2pd & 166x166 & 1660 & -1.17230096 & -1.16788170 & -1.16792099 & 0.025 & 2.748 \\
aug-cc-PVTZ & 350x350 & 3500 & -1.17258553 & -1.16637818 & -1.16637972 & 0.001 & 3.894 \\
cc-PVQZ & 552x552 & 5520 & -1.17375026 & -1.16599589 & -1.16604277 & 0.029 & 4.836 \\
def2-QZVP & 552x552 & 5520 & -1.17382320 & -1.16645322 & -1.16653460 & 0.051 & 4.574 \\
aug-cc-PVQZ & 1256x1256 & 12560 & -1.17382181 & -1.16357644 & No data\footnotemark[4] & - & -

\end{tabular}
\end{ruledtabular}
\footnotetext[1]{Energy obtained using the reference diagonalization.}
\footnotetext[2]{Energy obtained using the QAE in the classical mode (Tabu search).}
\footnotetext[3]{Energy obtained using the QAE in the hardware mode (D-Wave 2000Q).}
\footnotetext[4]{The QAE failed in the hardware mode (see text).}
\end{table*}

\begin{table*}
\fontsize{7}{8.4}\selectfont
\caption{\label{table2} Absolute electronic ground state energies ($E_\textrm{h}$) and energy differences (kcal/mol). }
\begin{ruledtabular}
\begin{tabular}{cccccccccc}

Molecule & Method & Basis & Mat. size & QUBO size & $E_{ref}$\footnotemark[1] & $E_{cl}$\footnotemark[2] & $E_{hw}$\footnotemark[3] & $E_{cl}-E_{hw}$ & $E_{hw}-E_{ref}$ \\

\hline
H$_2$       & FCI           & STO-3G     & 2x2       & 20    &  -1.87980819 &  -1.87980819 &  -1.87980819 &  0.000 & 0.000 \\
HF          & FCI           & STO-3G     & 18x18     & 180   & -98.60174790 & -98.60150630 & -98.60140647 & -0.063 & 0.214 \\
H$_2$O      & FCI           & STO-3G     & 133x133   & 1330  & -75.02039100 & -75.01165123 & -75.01178703 &  0.085 & 5.399 \\
H$_2$O      & CAS(8e,7o)SCF & cc-PVDZ    & 321x321   & 3210  & -76.11470570 & -76.10050279 & -76.10314815 &  1.660 & 7.252 \\
CH$_2^{2+}$ & FCI           & STO-3G     & 169x169   & 1690  & -37.40440097 & -37.39601005 & -37.39647755 &  0.293 & 4.972 \\
BeH$_2$     & FCI           & STO-3G     & 169x169   & 1690  & -15.59474568 & -15.59143293 & -15.59095276 & -0.301 & 2.380 \\
H$_3^+$     & FCI           & cc-PVTZ    & 532x532   & 5320  &  -1.34149794 &  -1.33044775 &  -1.33033386 & -0.071 & 7.005 \\
BH$_3$      & CAS(6e,6o)SCF & 6-311++G** & 208x208   & 2080  & -26.44129518 & -26.43066167 & -26.42990861 & -0.473 & 7.145 \\
BH$_3$      & FCI           & STO-3G     & 1250x1250 & 12500 & -26.12145752 & -26.10470471 & No data\footnotemark[5] & - & -

\end{tabular}
\end{ruledtabular}
\footnotetext[1]{Energy obtained using the reference diagonalization.}
\footnotetext[2]{Energy obtained using the QAE in the classical mode (Tabu search).}
\footnotetext[3]{Energy obtained using the QAE in the hardware mode (D-Wave 2000Q).}
\footnotetext[5]{The QAE failed in the hardware mode (see text).}
\end{table*}

\begin{table*}
\caption{\label{table3} Absolute electronic excited state energies ($E_\textrm{h}$) and energy differences (kcal/mol) of the H$_2$O molecule computed using the FCI/STO-3G matrix. }
\begin{ruledtabular}
\begin{tabular}{cccccc}

State \# & $E_{ref}$\footnotemark[1] & $E_{cl}$\footnotemark[2] & $E_{hw}$\footnotemark[3] & $E_{cl}-E_{hw}$ & $E_{hw}-E_{ref}$ \\

\hline

1 & -75.02039100 & -75.00825860 & -75.01096336 & 1.697 & 5.916 \\
2 & -74.53743573 & -74.52927677 & -74.52949083 & 0.134 & 4.985 \\
3 & -74.43197396 & -74.41290660 & -74.40564845 & -4.555 & 16.519 \\
4 & -74.31751275 & -74.31150179 & -74.30211833 & -5.888 & 9.660 \\
5 & -74.07950770 & -74.04154252 & -74.03510825 & -4.038 & 27.861 \\

\end{tabular}
\end{ruledtabular}
\footnotetext[1]{Energy obtained using the reference diagonalization.}
\footnotetext[2]{Energy obtained using the QAE in the classical mode (Tabu search).}
\footnotetext[3]{Energy obtained using the QAE in the hardware mode (D-Wave 2000Q).}
\end{table*}

% Rename figure caption labels
\makeatletter
\renewcommand{\fnum@figure}{FIG. ~S\thefigure}
\makeatother

\begin{figure}
\includegraphics[scale=0.7]{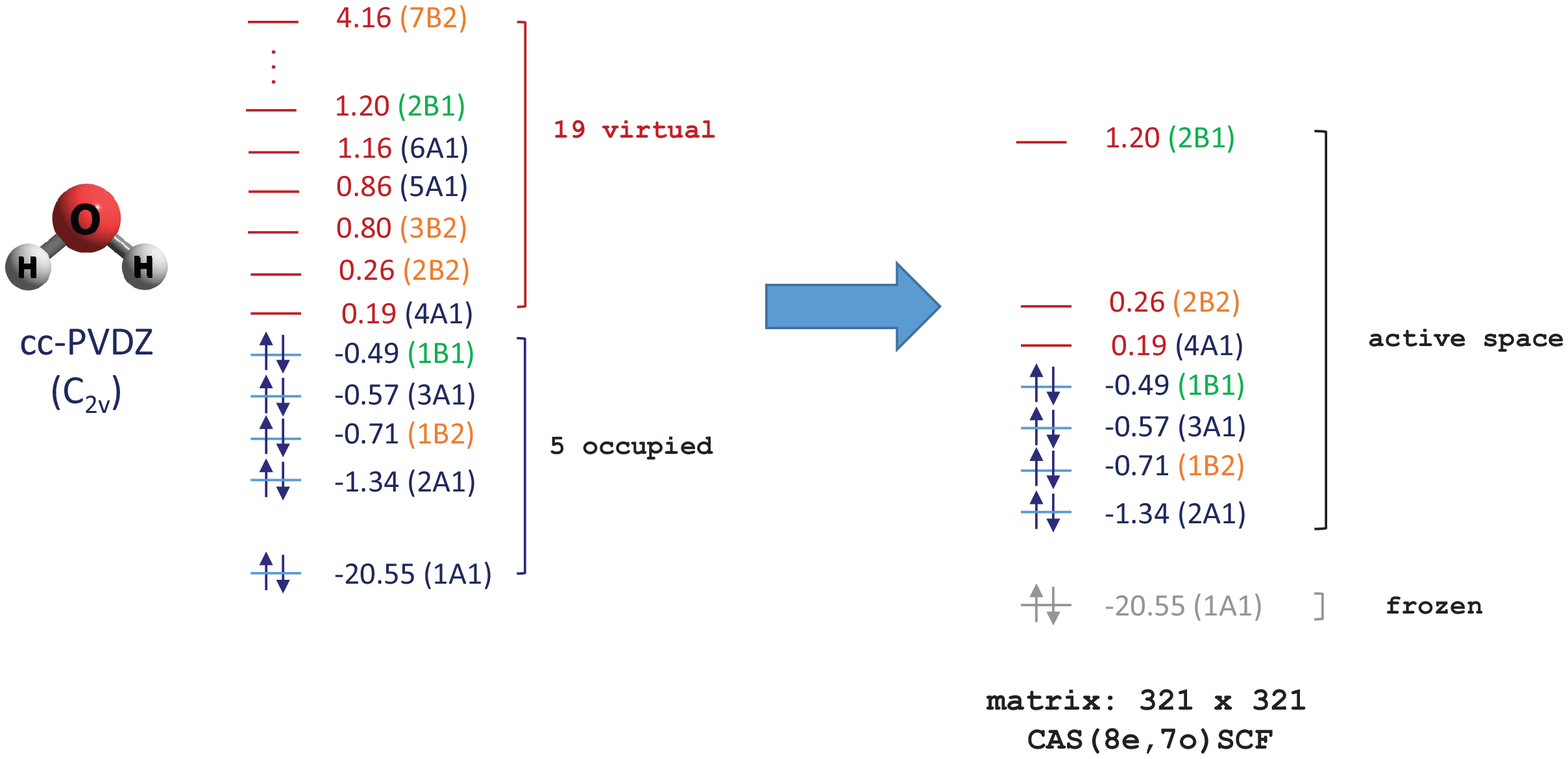}
\caption{\label{figS1} Active space selection for the H$_2$O molecule. The 2 lowest-energy electrons out of 10 were frozen (core) and 7 orbitals were selected out of 24. The resulting CAS(8e,7o)SCF calculation was done using the cc-PVDZ basis set.}
\end{figure}

\begin{figure}
\includegraphics[scale=0.7]{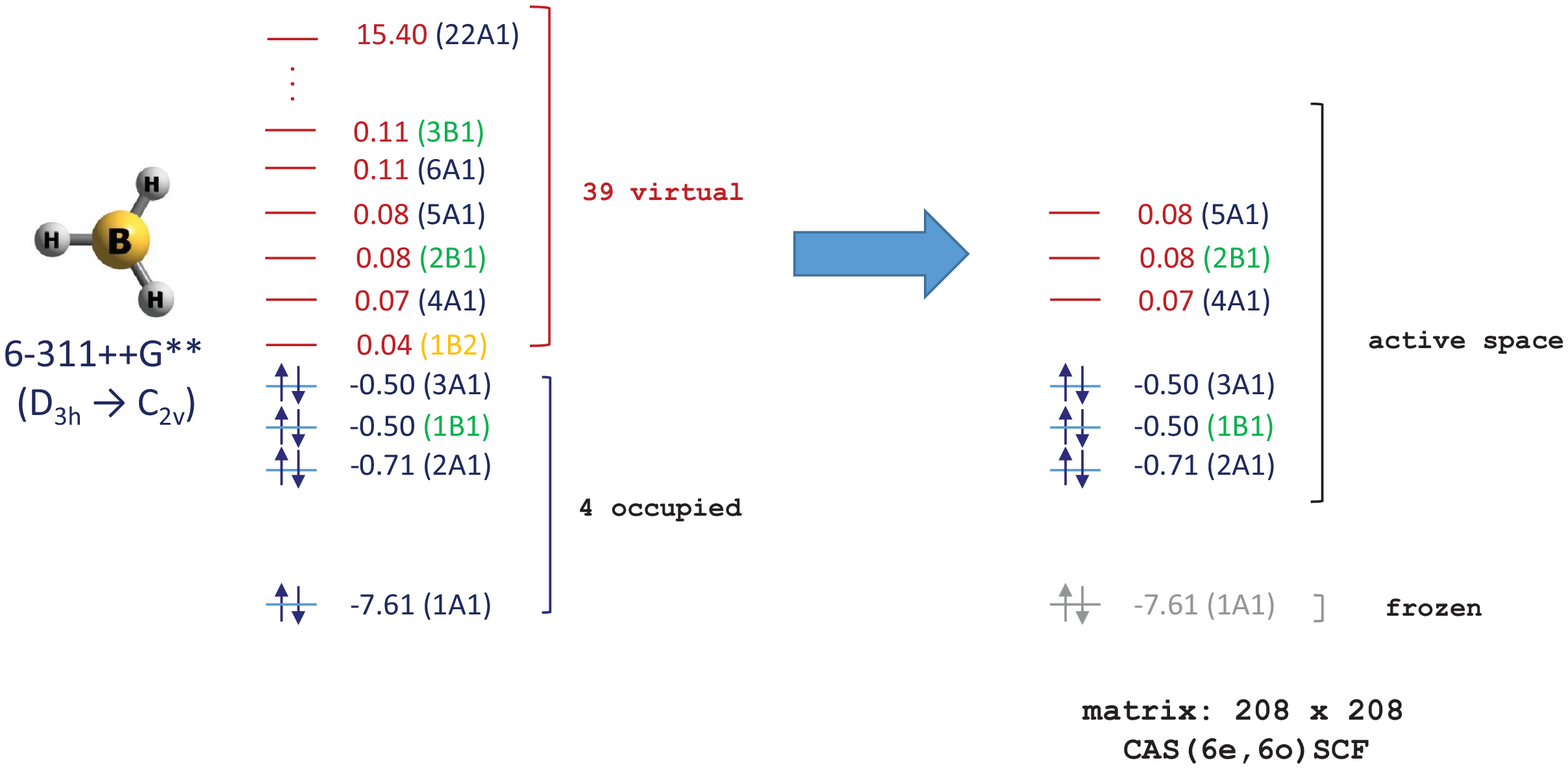}
\caption{\label{figS2} Active space selection for the BH$_3$ molecule. The 2 lowest-energy electrons out of 8 were frozen (core) and 6 orbitals were selected out of 43. The resulting CAS(6e,6o)SCF calculation was done using the 6-311++G** basis set.}
\end{figure}